\documentclass[letterpaper,12pt]{article}
\pdfoutput=1
\usepackage{jheppub}
\usepackage{epsfig}
\usepackage{amsmath}
\usepackage{subfigure}

\DeclareSymbolFont{AMSa}{U}{msa}{m}{n}
\DeclareSymbolFont{AMSb}{U}{msb}{m}{n}
\let\Box\relax
\DeclareMathSymbol{\Box}{\mathord}{AMSa}{"03}

\newcommand{\be}{\begin{equation}}
\newcommand{\ee}{\end{equation}}
\newcommand{\bea}{\begin{eqnarray}}
\newcommand{\eea}{\end{eqnarray}}

\newcommand{\f}{\frac}
\newcommand{\h}{\hspace*{1mm}}

\newcommand{\ds}{\displaystyle}
\newcommand{\ra}{\rightarrow}

\newcommand{\Od}{{\mathcal{O}}}
\newcommand{\Rd}{{\mathcal{R}}}

\title{On primordial equation of state transitions}

\author{Aditya Aravind,}
\author{Dustin Lorshbough}
\author{and Sonia Paban}
\affiliation[1]{Department of Physics and Texas Cosmology Center\\The University of Texas at Austin,
TX 78712.}
\emailAdd{Aditya.phy@utexas.edu}
\emailAdd{Lorsh@utexas.edu}
\emailAdd{Paban@zippy.ph.utexas.edu}

\abstract{We revisit the physics of transitions from a general equation of state parameter to the final stage of slow-roll inflation.  We show that it is unlikely for the modes comprising the cosmic microwave background to contain imprints from a pre-inflationary equation of state transition and still be consistent with observations.  We accomplish this by considering observational consistency bounds on the amplitude of excitations resulting from such a transition.  As a result, the physics which initially led to inflation likely cannot be probed with observations of the cosmic microwave background.  Furthermore, we show that it is unlikely that equation of state transitions may explain the observed low multipole power suppression anomaly.}

\preprint{$\begin{array}{c}\text{UTTG-03-16}\\\text{TCC-001-16}\end{array}$}

\begin{document}
\maketitle
\flushbottom
\section{Introduction}
The paradigm of inflation provides a suitable framework for understanding the observed spectrum of cosmological perturbations \cite{Ade:2015lrj}.  Many questions remain regarding the origin of inflation \cite{Yamauchi:2011qq,East:2015ggf,Kleban:2016sqm} and its duration \cite{Schwarz:2009sj,Cicoli:2014bja}.  It has been proposed that the transition to inflation may explain observed anomalies in the cosmic microwave background (CMB) if the duration of inflation is not too long \cite{Contaldi:2003zv}.

This paper revisits the generation of spectrum excitations due to the transition from an arbitrary equation of state parameter $(w)$ to inflation.  We emphasize the importance of using the proper matching conditions across the transition and show that previous studies have drawn incorrect conclusions when using improper matching conditions.  By combining observational consistency bounds on the excitation amplitude with the proper matching conditions we show that the cosmic microwave background likely does not contain imprints from the pre-inflationary universe.  Our study emphasizes three points:\\
\indent 1.  Observation strongly bounds the amplitude of excitation.\\
\indent 2.  The fractional change in $w$ must be small if a $w$ transition is observed.\\
\indent 3.  It is unlikely that transitions explain the $(20\lesssim\l\lesssim30)$ anomaly.\\
The first point has been made before \cite{Greene:2004np,Aravind:2013lra,Flauger:2013hra,Aravind:2014axa}, but a more general formulation of the bound is presented in section \ref{sec:Single}.  In section \ref{sec:Instant} we analytically compute the excited spectrum for the case of an instant transition, allowing us to bound the pre-transition equation of state parameter.  In section \ref{sec:Gradual} we discuss the excited spectrum for the case of a gradual transition which is modeled by a hyperbolic tangent function.  In section \ref{sec:low} we address the low power anomaly occurring for multipoles $20\lesssim\l\lesssim30$ and show that the type of models we discuss cannot explain the anomaly.  In section \ref{sec:Conclusions}, we conclude.

\section{Single Field Inflation}\label{sec:Single}
\subsection{Overview}
In this section we will review the basic equations of inflationary theory, emphasizing the appearance of $\dot{\epsilon}$ terms which will play an important role in studying the enhanced spectrum that results from equation of state transitions.  The simplest theory for inflation that is consistent with observational data is a minimally coupled scalar field with an FRW metric \cite{Malik:2008im},
\begin{equation}\label{eq:setup}
S=\int \sqrt{-g}\left[\f{1}{2}M_P^2R-\f{1}{2}(\partial\phi)^2-V\right],\h\h\h ds^2=-dt^2+a^2dx^2.
\end{equation}
The Einstein field equations may be combined to obtain an expression relating the rate of Hubble parameter change directly to the sum of the energy density and the pressure of the contents in the universe,
\begin{equation}\label{eq:EFE}
2M_P^2\dot{H}=-(\rho+P).
\end{equation}
For a single scalar field with negligible gradients we may write $\rho=K+V$ and $P=K-V$.  It is convenient to introduce three dimensionless parameters,
\begin{equation}\label{eq:epsilon}
\epsilon_H=-\f{\dot{H}}{H^2},\h\h\h\epsilon_\phi=3\f{K_\phi}{\rho}\text{ and }w=\f{P}{\rho}=-1+\f{2}{3}\epsilon.
\end{equation}
It is clear from \eqref{eq:EFE} that $\epsilon_H=\epsilon_\phi$.  In order to obtain an accelerating geometry we require that the parameter $\epsilon_H$ be smaller than unity,
\begin{equation}\label{eq:accel}
\ddot{a}=a H^2(1-\epsilon_H)>0\text{ if }\epsilon_H<1.
\end{equation}

Cosmological observables are obtained by computing correlation functions of gauge invariant fluctuations.  The action for the comoving curvature perturbation is given by
\begin{equation}
S_\Rd=-M_P^2\int \sqrt{-g}\h\epsilon\h(\partial\Rd)^2.
\end{equation}
The fields may be Fourier decomposed as
\begin{equation}\label{eq:flds}
\hat{\Rd}=\int\h\f{d^3k}{(2\pi)^3}\h\hat{\Rd}_{\vec{k}}\h e^{i\vec{k}\cdot\vec{x}}, \h\h\h\hat{\Rd}_{\vec{k}}=\Rd_k\h \hat{a}_{\vec{k}}^\dagger+\Rd_k^*\h \hat{a}_{-\vec{k}},\h\h\h [\hat{a}_k,\hat{a}_{k'}^\dagger]=(2\h\pi)^3\delta^3\left(\vec{k}-\vec{k}'\right).
\end{equation}
The resulting equation of motion for the mode function is\footnote{The absence of a friction term proportional to $\dot{\epsilon}$ in the equation of motion for the tensor mode functions is the reason tensor modes are not enhanced for equation of state transitions in the way that scalars are enhanced.}
\begin{equation}\label{eq:eom}
\ddot{\Rd}_k+\left(3H+\f{\dot{\epsilon}}{\epsilon}\right)\dot{\Rd}_k+\f{k^2}{a^2}\Rd_k=0.
\end{equation}
For the case of quasi-de Sitter expansion ($w\approx-1$, $\epsilon\ll1$), the solution of lowest energy density is given by the Bunch-Davies solution
\begin{equation}\label{eq:mode_BD}
\Rd_{k,BD}=\f{1}{\sqrt{2\h\epsilon}}\f{H}{M_P\sqrt{2\h k^3}}\h\left(1+i\f{k}{a\h H}\right)e^{-ik/aH}.
\end{equation}

The scalar power spectrum has been measured to great accuracy, while the scalar bispectrum and tensor power spectrum have not yet been detected.  The scalar power spectrum is typically parameterized by an amplitude $A_S$ and a scale dependence $n_s$,
\begin{equation}
\langle\hat{\Rd}_{\vec{k}}\hat{\Rd}_{\vec{k}'}\rangle=(2\pi)^3\delta^3\left(\vec{k}+\vec{k}'\right)P_\Rd(k),\h\h\h\Delta_\Rd^2(k)=\f{k^3}{2\pi^2}P_\Rd(k)=A_S\left(\f{k}{k_*}\right)^{n_s-1+\Od(dn_s/dk)}.
\end{equation}
The observationally obtained values for $A_S$ and $n_s$ are given in Table \ref{tab:values}.

\begin{table}[h]
\centering
\begin{tabular}{|c|c|}
\hline
Scalar Power Spectrum Amplitude&Scalar Power Spectrum Tilt\\\hline
$\ln\left(10^{10}A_S\right)=3.094\pm0.034\h(1\sigma)$&$n_s=0.9645\pm0.0049\h(1\sigma)$\\\hline
\end{tabular}
\caption{Cosmological parameters as measured by Planck TT+TE+EE+lowP \cite{Ade:2015lrj,Adam:2015rua,Ade:2015ava,Planck:2015xua}.  The pivot scale used by the Planck experiment is $k_*=0.05$ Mpc$^{-1}$.}
\label{tab:values}
\end{table}

The predictions for the Bunch-Davies mode functions are
\begin{equation}
A_S=\f{1}{8\pi^2\epsilon}\f{H^2}{M_P^2},\h\h\h n_s=1+2\eta-4\epsilon,\h\h\h\eta=-\f{\ddot{\phi}}{H\dot{\phi}},
\end{equation}

\subsection{Observables for General Bogoliubov Parameters}
The Bunch-Davies state is the solution of lowest energy density in quasi-de Sitter expansion.  However, in general the solution does not need to be the solution of lowest energy density.  We write excited solutions as a Bogoliubov transformation of the Bunch-Davies solution
\begin{equation}
\Rd_{k,\text{excited}}=\alpha_k\h\Rd_{k,BD}+\beta_k\h\Rd_{k,BD}^*\text{ and }\hat{a}_{\vec{k},\text{excited}}^\dagger=\alpha_k^*\h\hat{a}_{\vec{k},BD}^\dagger-\beta_k^*\h\hat{a}_{-\vec{k},BD}.
\end{equation}
Satisfying the canonical commutation relations requires that $|\alpha_k|^2-|\beta_k|^2=1$.

Excited states change the cosmological parameters.  In terms of the expressions corresponding to the Bunch-Davies solution previously discussed, the new parameter expressions are given by
\begin{equation}\label{eq:ab_mod_scalar}
\Delta_{\Rd,\text{excited}}^2=\Delta_{\Rd,\text{BD}}^2|\alpha_k+\beta_k|^2,\h\h\h A_S=A_{S,BD}|\alpha_{k_*}+\beta_{k_*}|^2,\h\h\h n_s=n_{s,BD}+\f{d\ln|\alpha_k+\beta_k|^2}{d\ln k}.
\end{equation}

\subsection{Bounds on Excitation Amplitude}
We present the bounds arising from backreaction considerations for different functional forms of Bogoliubov excitations, extending the results of \cite{Aravind:2013lra,Flauger:2013hra,Aravind:2014axa} to other functional forms, to show that the limits are similar.  Bounds that arise from measurements of $n_s$ and the observational limits on $f_{\text{NL}}^{\text{loc}}$ turn out to be weaker than the bounds that are obtained from backreaction considerations.

\subsubsection{Bounds from Backreaction Considerations}
The scalar power spectrum has been measured to deviate only slightly from scale invariance.  This implies that the modes comprising the observable cosmic microwave background should not vary dramatically in amplitude across the approximately 3-4 decades of modes we observe today.  Therefore the modes we observe today should either be excited modes or Bunch-Davies modes.  We will compute the bounds on excitation parameters implied by this.

In order for all of the modes comprising the CMB to have exited the horizon during inflation, the highest-$\l$ mode must be at least $n\approx3-4$ \cite{Peiris:2009wp,Hunt:2013bha,Hunt:2015iua} decades shorter wavelength than the horizon size at the beginning of inflation.  To obtain the most conservative bound from backreaction considerations we will assume that modes with higher momentum than the observable CMB modes are not excited.  The highest momentum excited mode therefore satisfies
\begin{equation}\label{eq:ratio}
p_{\text{UV}}\geq10^n H.
\end{equation}

The energy density stored in the fluctuations after adiabatic subtraction is given by
\begin{equation}
\langle\rho_\Rd\rangle=M_P^2\epsilon\int_0^{\infty}\f{d^3k}{(2\pi)^3}\left[|\dot{\Rd}_{k,\text{excited}}|^2-|\dot{\Rd}_{k,BD}|^2+\f{k^2}{a^2}\left(|\Rd_{k,\text{excited}}|^2-|\Rd_{k,BD}|^2\right)\right],
\end{equation}
We will consider different functional forms of scale dependence for the Bogoliubov parameter $\beta_k=|\beta|f_k$ shown in Table \ref{tab:f_k}.

\begin{table}[h]
\centering
\begin{tabular}{|c|c|}\hline
$f_k$&$\langle\rho_\Rd\rangle$\\\hline
$\Theta\left(k_{\text{UV}}-k\right)$&$\ds{\f{|\beta|^2}{8\pi^2}\left(p_{\text{UV}}^4+H^2p_{\text{UV}}^2\right)}$\\
&\\
$\exp[-k/k_{\text{UV}}]$&$\ds{\f{|\beta|^2}{16\pi^2}\left(3p_{\text{UV}}^4+H^2p_{\text{UV}}^2\right)}$\\
&\\
$\exp[-(k/k_{\text{UV}})^2]$&$\ds{\f{|\beta|^2}{16\pi^2}\left(p_{\text{UV}}^4+H^2p_{\text{UV}}^2\right)}$\\\hline
\end{tabular}
\caption{The energy density for the different functional forms of $\beta_k=|\beta|f_k$ that are considered.  The first form is that used in \cite{Greene:2004np,Aravind:2013lra,Flauger:2013hra,Aravind:2014axa}.  As emphasized in \cite{Carney:2011hz}, the oscillations resulting from instant transitions are very rapid so the effective $\beta_k$ seen by experiments would appear nearly scale invariant. Bounds on the gaussian form were considered in \cite{Holman:2007na,Ganc:2011dy,Ashoorioon:2013eia,Ashoorioon:2014nta}.  For a discussion on coherent states see \cite{Kundu:2011sg,Kundu:2013gha}.}
\label{tab:f_k}
\end{table}

The energy density stored in the fluctuations should remain sub-dominant to the kinetic energy of the inflaton field, $\epsilon M_P^2H^2$, in order for $\epsilon_H=\epsilon_\phi$ to remain true.  This backreaction bound may be written as
\begin{equation}
\langle\rho_\Rd\rangle<\epsilon M_P^2H^2=\f{H^4}{8\pi^2 A_S}|\alpha_{k_*}+\beta_{k_*}|^2.
\end{equation}
It is convenient to introduce a parameter which is the coefficient ratio for the leading energy density terms in Table \ref{tab:f_k},
\begin{equation}
c_f=\left\{
\begin{array}{ll}
1,&f_k=\Theta\left(k_{\text{UV}}-k\right)\\
2/3,&f_k=\exp[-k/k_{\text{UV}}]\\
2,&f_k=\exp[-(k/k_{\text{UV}})^2]
\end{array}
\right..
\end{equation}
The backreaction bound gives us,
\begin{equation}\label{eq:pUV_UB}
p_{\text{UV}}\lesssim\f{H|\alpha_{k_*}+\beta_{k_*}|^{1/2}}{A_S^{1/4}|\beta|^{1/2}}\times c_f^{1/4}
\end{equation}

Recalling that $|\alpha_k|^2=1+|\beta_k|^2$, using \eqref{eq:ratio} and \eqref{eq:pUV_UB} we obtain an upper bound for $|\beta|$ given by
\begin{equation}
|\beta|\lesssim10^{-2n}A_S^{-1/2}c_f^{1/2}.
\end{equation}
We provide the numerical evaluation of the $|\beta|$ upper bound and the corresponding bounds on $|\alpha_k+\beta_k|$ in Table \ref{tab:instant}.

\begin{table}[h]
\centering
\begin{tabular}{|c|c|c|c|}\hline
$f_k$&$|\beta|$ UB&$|\alpha_{k_*}+\beta_{k_*}|$ LB&$|\alpha_{k_*}+\beta_{k_*}|$ UB\\\hline
$\Theta\left(k_{\text{UV}}-k\right)$&0.022&0.97&1.022\\
$\exp[-k/k_{\text{UV}}]$&0.018&0.98&1.02\\
$\exp[-(k/k_{\text{UV}})^2]$&0.031&0.97&1.03\\\hline
\end{tabular}
\caption{Bounds on $\beta_k=|\beta|f_k$ and $|\alpha_k+\beta_k|$ arising from backreaction considerations.  We have taken $n=3$ and $k_*=0.05$ Mpc$^{-1}\sim0.5k_{\text{UV}}$ for the numerical estimates.  To obtain the lower bound we have allowed $\beta_k$ to be negative.}
\label{tab:instant}
\end{table}

\subsubsection{Special Case: Only $\l\lesssim30$ Modes Excited}\label{sec:special}
For the lowest $\l$ cosmic variance limited modes ($\l\lesssim30$), one may not use scale invariance to bound the excitation amplitude.  Instead, we compare the theoretical prediction with the observational error bar width \cite{Adam:2015rua} to obtain a conservative estimate of $0.5\lesssim|\alpha_k+\beta_k|^2\lesssim 2$.

\section{Excitation Mechanism: Instant Transition}\label{sec:Instant}
\begin{figure}[h]
\centering
\includegraphics[width=12cm,height=8cm]{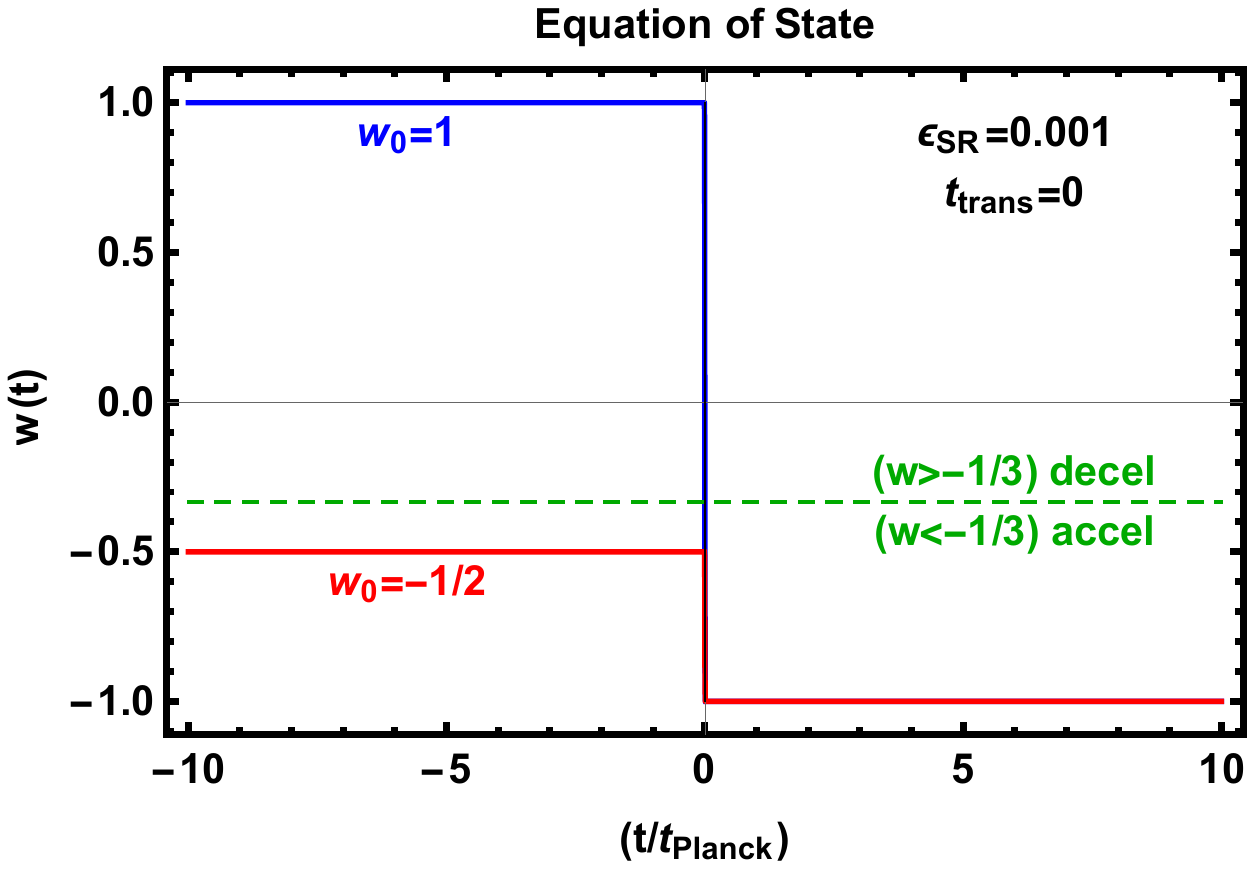}
\caption{Examples of instantaneous transitions.  Though not explicitly plotted, we also have studied the transition which increases the equation of state parameter (see Table \ref{tab:largest_w0_2}).}
\label{Fig:instantw}
\end{figure}

Although unrealistic, the idealized case of an instantaneous transition from one equation of state parameter value to a different value is useful since it allows for an analytical calculation of the excited spectrum.  In the next section, the more realistic case of a gradual transition will be discussed.  Figure \ref{Fig:instantw} illustrates the transition for several different values of $w_0$.

\subsection{Matching Conditions}
The instant transitions that we are considering effectively have a discontinuity in the slow roll parameter $\epsilon$.  Therefore, one must be careful when determining which quantities related to $\Rd$ are continuous since the evolution of $\Rd$ depends explicitly on $\dot{\epsilon}$ according to \eqref{eq:eom}.  It was emphasized in \cite{Carney:2011hz,Deruelle:1995kd} that the proper matching conditions are given by
\begin{equation}\label{eq:matching}
\left[\Rd_k\right]_\pm=0,\h\h\h\left[\epsilon\dot{\Rd}_k\right]_\pm=0,
\end{equation}
where we have used the notation that $\left[\cdots\right]_\pm$ denotes the change in a quantity across the transition.  The origin of these conditions may be easily seen from the differential equation for scalar fluctuations \eqref{eq:eom}, which can be rewritten as
\begin{equation}
\f{d}{dt}\left(a^3\epsilon\dot{\Rd}_k\right)=-k^2a\epsilon\Rd_k.
\end{equation}
The fluctuation mode function itself, $\Rd_k$, is continuous.  If we note that $\epsilon=d(1/H)/dt$, we may time integrate both sides of the equation close to the transition to obtain the continuity condition on the derivative of the mode function,
\begin{equation}
\left[a^3\epsilon\dot{\Rd}_k\right]_\pm=\lim_{\delta\ra0}\int_{(\Delta t)_{t}=-\delta}^{(\Delta t)_t=\delta} dt\left(-k^2a\epsilon\Rd_k\right)=\left[-k^2\f{a}{H}\Rd_k\right]_\pm=0.
\end{equation}
We have introduced the notation $(\Delta t)_t=\left(t-t_{\text{trans}}\right)$ to denote that time difference between the cosmic time and the time of transition.

Microscopically, the inflaton field will take on a uniform value\footnote{Note that $\delta\phi=0$ in the comoving gauge \cite{Malik:2008im}.} at the transition time $(\Delta t)_t=0$.  This implies in the language of \cite{Deruelle:1995kd,Carney:2011hz} that the transition is characterized by a spacetime hypersurface of constant field value, which directly yields \eqref{eq:matching}.  This is in contrast to the examples studied in \cite{Deruelle:1995kd} in which the transition was characterized by a surface of constant energy density and hence the uniform density gauge is more appropriate since $\delta\rho=0$ in that gauge.  The continuity conditions \eqref{eq:matching} have been verified numerically by time evolving \eqref{eq:eom}.

\subsection{Observables: Allowed Parameter Space}
Based on the matching conditions previously discussed, we would like to solve the following system of equations:
\begin{equation}\label{eq:system}
\begin{array}{lccc}
\ds{(\Delta t)_t=0:}&\ds{\alpha_{0,k}\Rd_{0,k}+\beta_{0,k}\Rd_{0,k}^*}&\ds{=}&\ds{\alpha_{f,k}\Rd_{\text{SR},k}+\beta_{f,k}\Rd_{\text{SR},k}^*,}\\
&&&\\
\ds{(\Delta t)_t=0:}&\ds{\epsilon_0\left(\alpha_{0,k}\dot{\Rd}_{0,k}+\beta_{0,k}\dot{\Rd}_{0,k}^*\right)}&\ds{=}&\ds{\epsilon_{\text{SR}}\left(\alpha_{f,k}\dot{\Rd}_{\text{SR},k}+\beta_{f,k}\dot{\Rd}_{\text{SR},k}^*\right).}
\end{array}
\end{equation}
The explicit forms of the functions $\Rd_{0,k}$ and $\Rd_{\text{SR},k}$ are given in \eqref{eq:mode_fns}.  The coefficients $\left\{\alpha_{0,k},\beta_{0,k}\right\}$ account for the fact that the spectrum may be excited prior to the transition to slow-roll inflation.  The case of the lowest energy density vacuum state transitioning to inflation is given by the choice $\alpha_{0,k}=1$ and $\beta_{0,k}=0$.

In order to solve \eqref{eq:system}, we must find a solution to scalar fluctuation mode equation which is properly normalized.  The normalization condition is given by the canonical commutation relation, which may be rewritten as a condition on the Wronskian of the scalar fluctuation mode function as follows
\begin{equation}
2a^3M_P^2\epsilon\left(\dot{\Rd}_k\Rd^*_k-\Rd_k\dot{\Rd}_k^*\right)=i.
\end{equation}
The background geometry evolution is given by
\begin{equation}
\epsilon=\left\{\begin{array}{ll}
\epsilon_{\text{SR}}&(\Delta t)_t<0\\
\epsilon_0&(\Delta t)_t>0
\end{array}
\right.,\h\h\h H(t)=H_t\left[1+\epsilon\h H_t(\Delta t)_t\right]^{-1}
,\h\h\h a(t)=a_t\left(\f{H_t}{H(t)}\right)^{1/\epsilon}.
\end{equation}
We have defined $H_t$ and $a_t$ as the Hubble parameter and scale factor at the time of transition, $(\Delta t)_t=0$.  It is convenient to introduce the variables
\begin{equation}
\begin{array}{lc}
\ds{(\Delta t)_t<0:}&\ds{\widetilde{\tau}_0=\int\f{dt}{a}=\f{1}{(-1+\epsilon_0)}\f{1}{a(t)H(t)},}\\
&\\
\ds{(\Delta t)_t>0:}&\ds{\widetilde{\tau}_{\text{SR}}=\int\f{dt}{a}=\f{1}{(-1+\epsilon_{\text{SR}})}\f{1}{a(t)H(t)}.}
\end{array}
\end{equation}
The properly normalized solution to the scalar equation of motion \eqref{eq:eom} for a constant equation of state is given as\footnote{Note that $H_\nu^{1,2}(x\ra\infty)\ra e^{\pm ix}$.}
\begin{equation}\label{eq:mode_fns}
\begin{array}{lcc}
\ds{(\Delta t)_t<0:}&\ds{\Rd_{0,k}(t)=\sqrt{\f{\pi}{2^3\h\epsilon_0}}\f{1}{M_P\h a(t)}\sqrt{|\widetilde{\tau}_0|}\h H_{\nu_0}^2\left[k|\widetilde{\tau}_0|\right],}&\ds{\nu_0=\f{3}{2}\f{(1-w_0)}{(1+3w_0)}}\\
&&\\
\ds{(\Delta t)_t>0:}&\ds{\Rd_{\text{SR},k}(t)=\sqrt{\f{\pi}{2^3\h\epsilon_{\text{SR}}}}\f{1}{M_P\h a(t)}\sqrt{|\widetilde{\tau}_{\text{SR}}|}\h H_{\nu_{\text{SR}}}^2\left[k|\widetilde{\tau}_{\text{SR}}|\right],}&\ds{\nu_{\text{SR}}=\f{3}{2}\f{(1-w_{\text{SR}})}{(1+3w_{\text{SR}})}}
\end{array}
\end{equation}

Employing the matching conditions \eqref{eq:system} we are able to write the Bogoliubov parameters for an arbitrary choice of $w_0$ and initial excitation parameters $\alpha_0$ and $\beta_0$ as
\begin{equation}\label{eq:coeffs}
\alpha_{f,k}=\sqrt{\f{\epsilon_0}{\epsilon_{\text{SR}}}}\sqrt{\left|\f{\epsilon_{\text{SR}}-1}{\epsilon_0-1}\right|}\f{\alpha_{\text{num}}}{\alpha_{\text{denom}}},\h\h\h \beta_{f,k}=\alpha_{f,k}\text{ with }H^{1}_{f(\nu_{\text{SR}})}\leftrightarrow H^{2}_{f(\nu_{\text{SR}})},
\end{equation}
\begin{equation}
\begin{array}{l}
\alpha_{\text{num}}=\\
\left(\begin{array}{c}
s_{\tau_0}H^1_{\nu_{\text{SR}}}(x_{\text{SR}})\left\{\alpha_0\left[H^2_{\nu_0-1}(x_0)-H^2_{\nu_0+1}(x_0)\right]+\beta_0\left[H^1_{\nu_0-1}(x_0)-H^1_{\nu_0+1}(x_0)\right]\right\}\\
+\left[\alpha_0H^2_{\nu_0}(x_0)+\beta_0H^1_{\nu_0}(x_0)\right]\left\{\f{\epsilon_{\text{SR}}}{\epsilon_0}\left[H^1_{\nu_{\text{SR}}-1}(x_{\text{SR}})-H^1_{\nu_{\text{SR}}+1}(x_{\text{SR}})\right]\right.\\
\left.+\f{a_tH_t}{k}H^1_{\nu_{\text{SR}}}(x_{\text{SR}})\left[-2\left(1-\f{\epsilon_{\text{SR}}}{\epsilon_0}\right)+s_{\tau_0}|\epsilon_0-1|+\f{\epsilon_{\text{SR}}}{\epsilon_0}|\epsilon_{\text{SR}}-1|\right]\right\}
\end{array}\right)\end{array},
\end{equation}
\begin{equation}
\alpha_{\text{denom}}=\left(\begin{array}{c}
\ds{H^1_{\nu_{\text{SR}}}(x_{\text{SR}})\left[H^2_{\nu_{\text{SR}}+1}(x_{\text{SR}})-H^2_{\nu_{\text{SR}}-1}(x_{\text{SR}})\right]}\\
\ds{-H^2_{\nu_{\text{SR}}}(x_{\text{SR}})\left[H^1_{\nu_{\text{SR}}+1}(x_{\text{SR}})-H^1_{\nu_{\text{SR}}-1}(x_{\text{SR}})\right]}
\end{array}\right).
\end{equation}
Here we have defined $x_0=|\epsilon_0-1|^{-1}(k/a_tH_t)$, $x_{\text{SR}}=|\epsilon_{\text{SR}}-1|^{-1}(k/a_tH_t)$ and $s_{\tau_0}=\text{sign}(\tilde{\tau}_0)$.

A special case of interest is a transition for which $\alpha_0=1$ and $\beta_0=0$, corresponding to a transition from the state of lowest energy density prior to the transition.  We will concentrate on this case for the remainder of the paper until section \ref{sec:low}.  The Bogoliubov parameters are oscillatory in nature, but in practice the oscillations can not be resolved experimentally and it is appropriate to approximate $\beta_{f,k}\approx\beta_f$.

To clarify what we mean when we state that the oscillations cannot be resolved experimentally, we compare the scale of oscillations to the binning scale used by the Planck experiment \cite{Planck:2015xua}.  The baseline Plik likelihood bin sizes are $\Delta\l=1$ for $\l<30$, $\Delta\l=5$ for $30\leq\l\leq99$, $\Delta\l=9$ for $100\leq\l\leq1503$, $\Delta\l=17$ for $1504\leq\l\leq2013$ and $\Delta\l=33$ for $2014\leq\l\leq2508$.  The oscillations in $\alpha_{f,k}$ and $\beta_{f,k}$ are controlled by $k\tau_{0,t}$, which results in several oscillations in a $\Delta\l=1$ window.

Having computed the excitation spectrum that results from an instant transition, we can translate the bounds obtained in the previous section into bounds on $w_0$.

\begin{figure}[h]
\centering
\includegraphics[width=7cm,height=7cm]{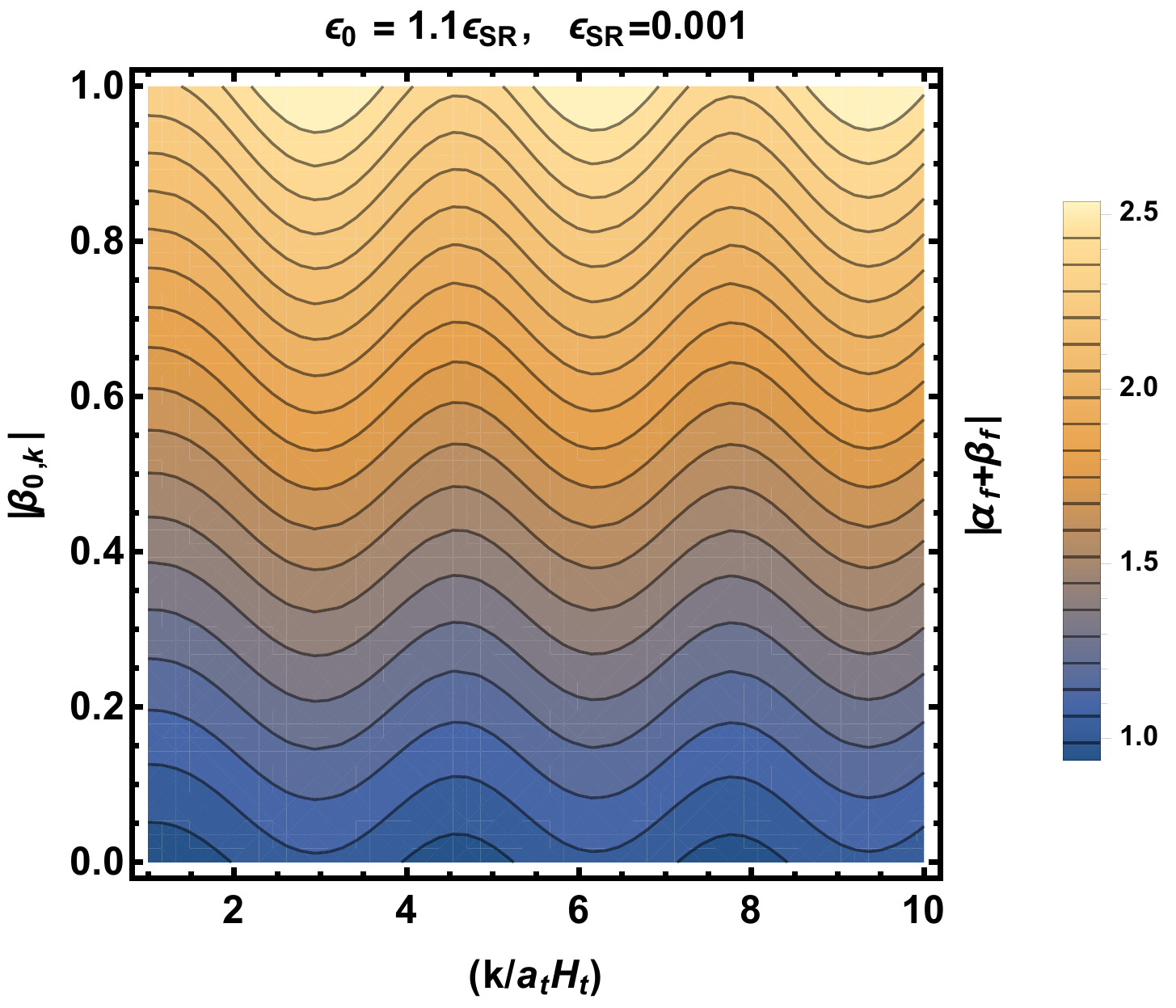}
\includegraphics[width=7cm,height=7cm]{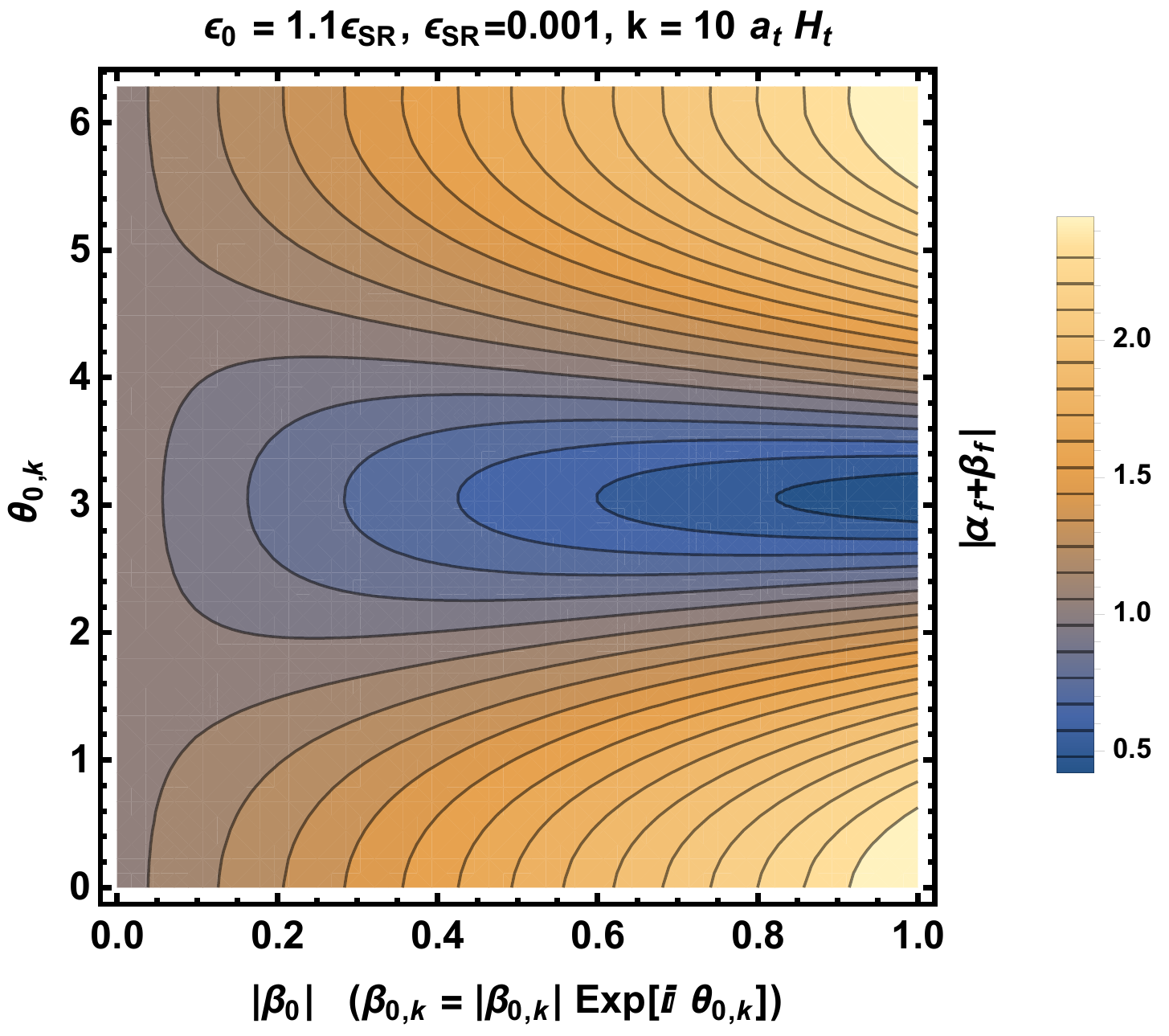}
\caption{Amplitude of excitations $|\alpha_{f,k}+\beta_{f,k}|$ after transition for initially excited state with $\beta_{0,k}\ne0$.  We have taken the form of the excitation prior to the transition to be $\beta_{0,k}=|\beta_{0,k}|\exp\left(i\theta_{0,k}\right)$ for arbitrary $\theta_{0,k}$ in the right plot and $\theta_{0,k}=0$ in the left plot.  These plots indicate that a larger $|\beta_{0,k}|$ typically leads to a larger deviation of $|\alpha_{f,k}+\beta_{f,k}|$ from unity.  This justifies our choice of $\beta_{0,k}=0$ to derive our bounds presented in Table \ref{tab:largest_w0}.}
\label{Fig:betavx}
\end{figure}

From our strongest bounds on $|\beta_f|$ summarized in Table \ref{tab:instant} and section \ref{sec:special}, we may tabulate the largest allowed $w_0$ for a given $\epsilon_{\text{SR}}$.  Consider the effect of $\beta_0=|\beta_0|\exp(i\theta_0)$.  We see from Figure \ref{Fig:betavx} that $|\alpha_f+\beta_f|$ is maximal for $\left\{|\beta_0|,\theta_0\right\}=\left\{\text{large},0\right\}$ and minimal for $\left\{|\beta_0|,\theta_0\right\}=\left\{\text{large},\pi\right\}$ for the parameter values specified on the plots.  Since we have identified both an upper and lower bound on $|\alpha_f+\beta_f|$, we will take the intermediate case of $\alpha_0=1$ and $\beta_0=0$.  In Table \ref{tab:largest_w0} we present the bounds on the fractional change of $w$ and $\epsilon$.

\begin{table}[h]
\centering
\begin{tabular}{|c|c|c|c|}\hline
Observed Multipoles Excited&Relevant Bound&$100|\f{w_0-w_{\text{SR}}}{w_{\text{SR}}}|$&$100|\f{\epsilon_0-\epsilon_{\text{SR}}}{\epsilon_{\text{SR}}}|$\\\hline
$\l\lesssim30$&$0.71\lesssim|\alpha_{f,k}+\beta_{f,k}|\lesssim1.41$&$\lesssim0.07$&$\lesssim94$\\
$\l>30$ and lower&$0.97<|\alpha_{f,k}+\beta_{f,k}|<1.022$&$\lesssim0.003$&$\lesssim4.3$\\\hline
\end{tabular}
\caption{The maximal allowed $\epsilon_0$ for a given $\epsilon_{\text{SR}}<\epsilon_0$.  We have reported the bounds for $\epsilon_{\text{SR}}=10^{-3}$, though we have confirmed that the bound on the fractional change of $\epsilon$ is not numerically sensitive to this input.  We have chosen $\alpha_0=1$ and $\beta_0=0$.}
\label{tab:largest_w0}
\end{table}

If the transition to inflation is well approximated as an instantaneous transition with an initial $w_0$ larger than is stated in Table \ref{tab:largest_w0}, the modes which are excited cannot comprise our observable CMB.  Note that modes which are super-Planckian at the time of transition should be described by the Bunch-Davies vacuum when their momenta redshift to become sub-Planckian in order for the stress-energy tensor to be renormalizable \cite{Carney:2011hz}.

We note for completeness that the bounds on the fractional change in $\epsilon$ are more restrictive if we consider a transition in which $\epsilon$ increases.  The bounds are explicitly given in Table \ref{tab:largest_w0_2}.  In Figure \ref{Fig:comparing_steps} we demonstrate how the morphology of the spectrum changes depending on whether the step in $\epsilon$ is to a smaller or larger $\epsilon$.

\begin{table}[h]
\centering
\begin{tabular}{|c|c|c|c|}\hline
Observed Multipoles Excited&Relevant Bound&$100|\f{w_0-w_{\text{SR}}}{w_{\text{SR}}}|$&$100|\f{\epsilon_0-\epsilon_{\text{SR}}}{\epsilon_{\text{SR}}}|$\\\hline
$\l\lesssim30$&$0.71\lesssim|\alpha_{f,k}+\beta_{f,k}|\lesssim1.41$&$\lesssim0.03$&$\lesssim43$\\
$\l>30$ and lower&$0.97<|\alpha_{f,k}+\beta_{f,k}|<1.022$&$\lesssim0.003$&$\lesssim3.3$\\\hline
\end{tabular}
\caption{The maximal allowed $\epsilon_0$ for a given $\epsilon_{\text{SR}}>\epsilon_0$.  We have reported the bounds for $\epsilon_{\text{SR}}=10^{-3}$, though we have confirmed that the bound on the fractional change of $\epsilon$ is not numerically sensitive to this input.  We have chosen $\alpha_0=1$ and $\beta_0=0$.}
\label{tab:largest_w0_2}
\end{table}

\subsection{Comparison with Previous Work}
There have been many previous studies analyzing equation of state transitions \cite{Carney:2011hz,Aravind:2013lra,Jain:2008dw,Chen:2015gla,Das:2014ffa,Cicoli:2014bja,Cai:2015nya,Contaldi:2003zv,Adshead:2011jq}.  Some recent examples with which we could easily compare our matching criteria are \cite{Chen:2015gla,Das:2014ffa,Cicoli:2014bja,Cai:2015nya,Contaldi:2003zv}.  The matching conditions used in those studies do not agree with the matching conditions presented in equation \eqref{eq:matching}.  We also note that studies which numerically evolve the Muhkanov-Sasaki equation without making approximations for $\left\{\dot{z},\ddot{z}\right\}$ should yield the correct result if a proper step size is chosen so that the transition is sampled.

One of our conclusions is that only transitions from one inflationary phase to another are allowed to be imprinted on the observable CMB.  A special case which has been studied previously is steps in the inflationary potential which are modeled by a hyperbolic tangent of the field value \cite{Adshead:2011jq}.  We find that even for the most violent case of an instant transition with a step size $|\Delta V|/V_0=\epsilon_0/3$, the fractional change in $\epsilon$ almost satisfies our least restrictive bound presented in Table \ref{tab:largest_w0_2}.  To see this explicitly, note that the initial kinetic energy is given by $K_0=|\Delta V|$ and therefore $K_f=2K_0$ by energy conservation.  The fractional change in $\epsilon\approx3K/V$ is given by
\begin{equation}
\epsilon_0=\f{1}{2}\f{\epsilon_{\text{SR}}}{(1+\epsilon_{\text{SR}}/6)}\approx\f{1}{2}\epsilon_{\text{SR}},\h\h\h |\f{\epsilon_0-\epsilon_{\text{SR}}}{\epsilon_{\text{SR}}}|\approx0.5.
\end{equation}

\section{Excitation Mechanism: Gradual Transition}\label{sec:Gradual}
\subsection{Transition Model}
\begin{figure}[h]
\centering
\includegraphics[width=12cm,height=8cm]{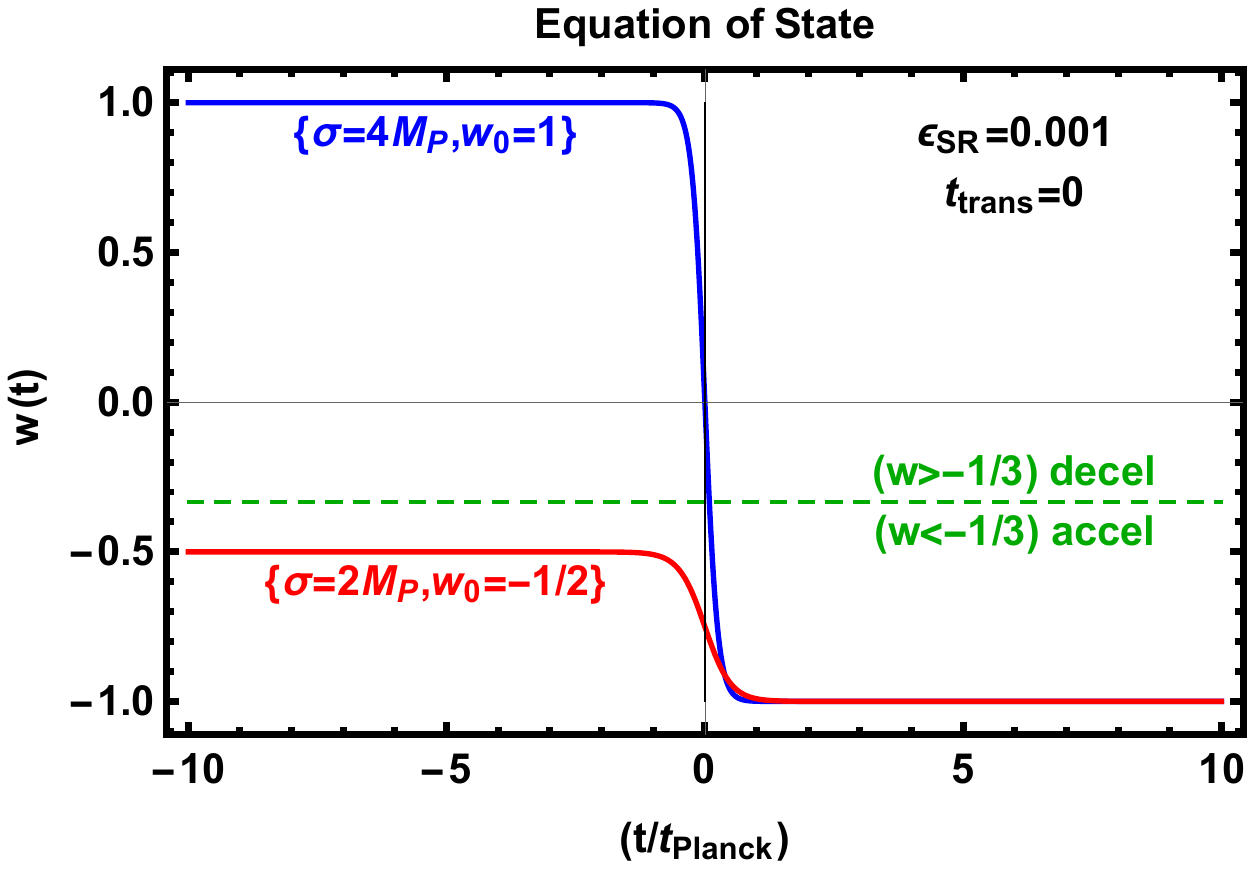}
\caption{The equation of state as a function of the cosmic time for gradual transitions parameterized by (\ref{eq:eqnw}).  Though not explicitly plotted, we also have studied transitions which increase the equation of state parameter (see Figure \ref{Fig:comparing_steps}).}
\label{fig:gradualw}
\end{figure}

Our parameterization is of the form
\begin{equation}\label{eq:eqnw}
w(t)=w_0+\f{1}{2}(w_{\text{SR}}-w_0)\left(1+\tanh\left[\sigma(\Delta t)_t\right]\right),\h w_{\text{SR}}=-1+\f{2}{3}\epsilon_{\text{SR}}.
\end{equation}
Figure \ref{fig:gradualw} illustrates the transition for two different values of $w_0$ and $\sigma$.  The slow-roll parameter is explicitly given by
\begin{equation}
\epsilon(t)=\f{3}{2}(1+w)=\f{1}{2}\left(\epsilon_{SR}+\epsilon_0+(\epsilon_{SR}-\epsilon_0)\tanh\left[\sigma(\Delta t)_t\right]\right).
\end{equation}

\subsection{Observables: Allowed Parameter Space}
\begin{figure}[h]
\centering
\includegraphics[width=12cm,height=8cm]{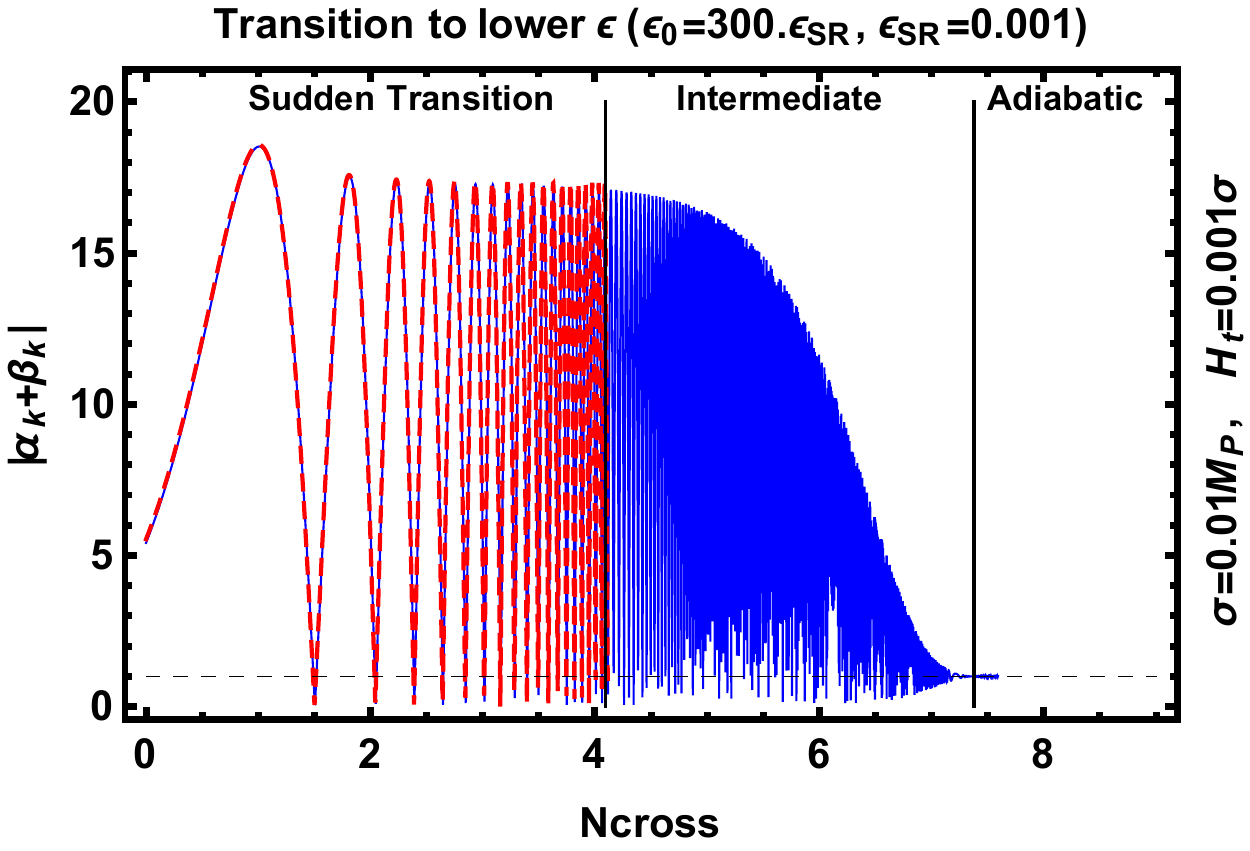}
\caption{A comparison of the spectrum enhancement obtained from the numerical solution (solid blue line) and the analytical formula (dashed red line) obtained from equation \eqref{eq:coeffs}.  The parameter $N_{\text{cross}}=\log\left(k/a_tH_t\right)$ is the number of efolds from the time of the transition until the mode exits the horizon.  For the sake of clarity, we end the analytically obtained line at 4 efolds.  The numerical solution stops at slightly over 7.5 efolds.  The separators between the three regimes of modes are approximately placed.  We have chosen $\sigma/H=10^3$ for illustration, though as $\sigma/H$ approaches unity fewer sub-horizon modes are excited.  For $\sigma/H\sim1$, there are still more than two efolds of excited modes which were sub-horizon at the time of transition.}
\label{fig:cases}
\end{figure}

\begin{figure}[h]
\centering
\includegraphics[width=7cm,height=5cm]{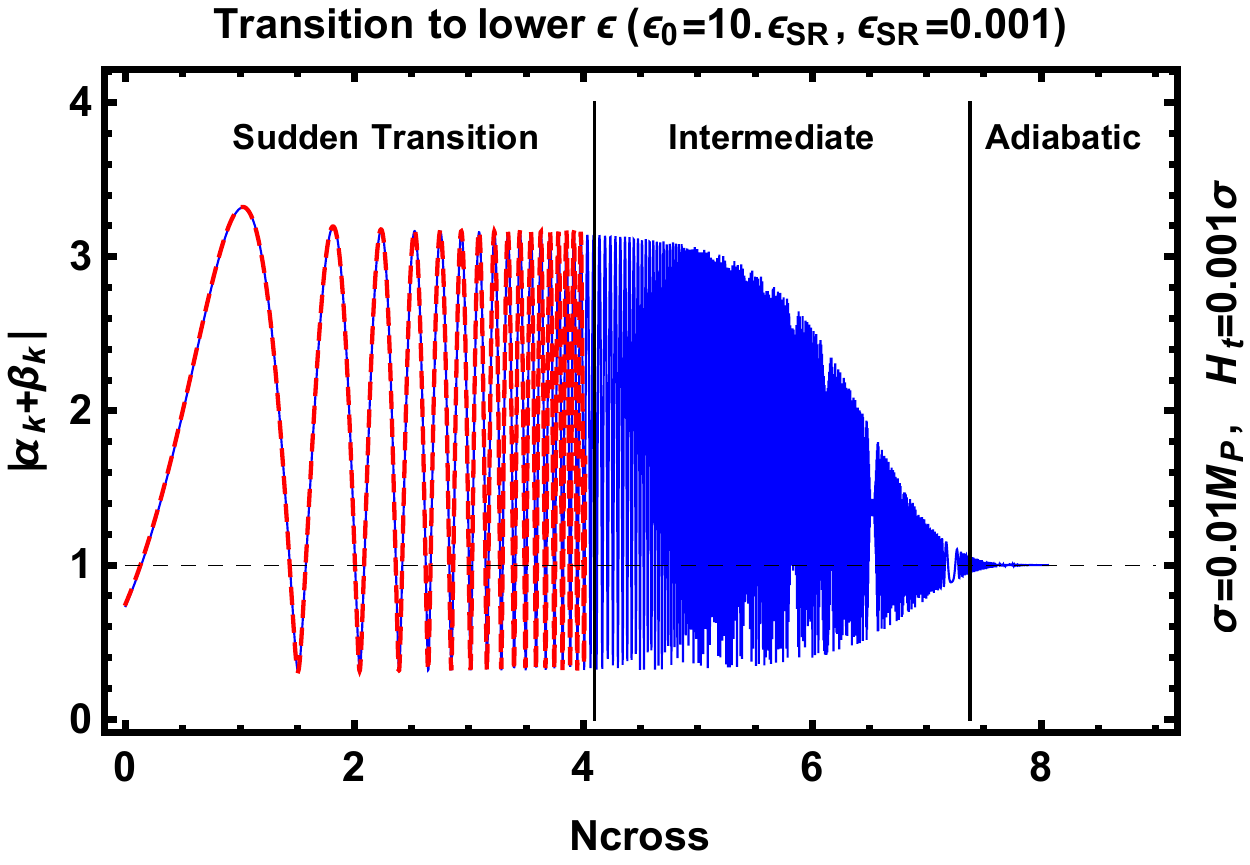}
\includegraphics[width=7cm,height=5cm]{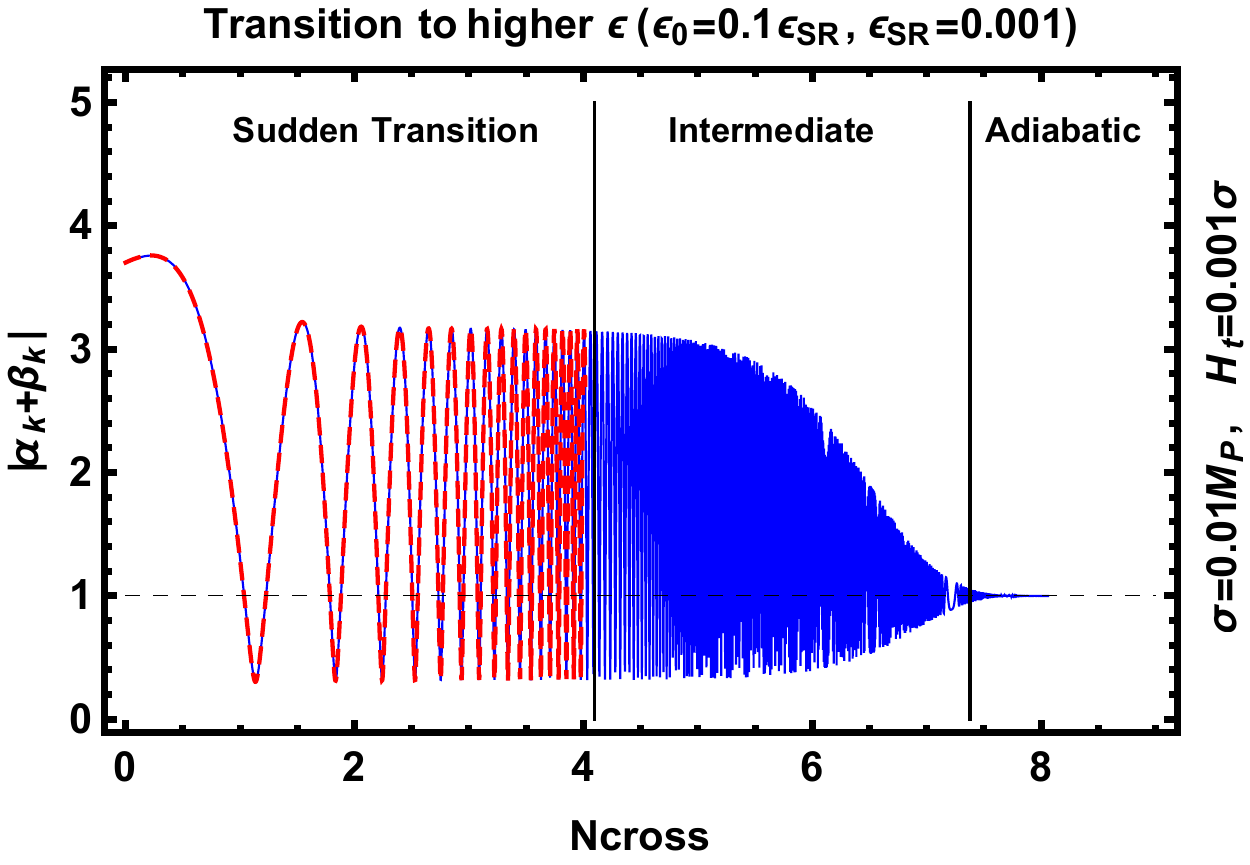}
\caption{Spectrum morphology comparison for a transition to a higher $\epsilon$ versus a transition to a lower $\epsilon$.  Note that a transition with $\epsilon_0>\epsilon_{\text{SR}}$ tends to reach its first peak for higher momenta than the case of $\epsilon_0<\epsilon_{\text{SR}}$.  The separators between the three regimes of modes are approximately placed.}
\label{Fig:comparing_steps}
\end{figure}

The gradual transition has three cases for modes depending on whether a mode experiences the transition as sudden, adiabatic or an intermediate case between the two.  For an example of these three regimes, see Figure \ref{fig:cases}.  We have explicitly compared the spectrum morphology for a case of transitioning to a lower $\epsilon$ to the case of transitioning to a higher $\epsilon$ in Figure \ref{Fig:comparing_steps}.

Comparing the $\dot{\epsilon}$ friction term and the frequency term in the equation of motion \eqref{eq:eom} provides an intuition for the three cases.  This is most easily done by rescaling the curvature perturbation in order to eliminate the friction term altogether.  The appropriate rescaling is given by
\begin{equation}
\Rd_k=(a^3\h \epsilon)^{-1/2}\h\bar{\Rd}_k.
\end{equation}
The resulting rescaled equation of motion is given by
\begin{equation}
\ddot{\bar{\Rd}}_k+\bar{\omega}_k^2\h\bar{\Rd}_k=0,\h\h\h\bar{\omega}_k^2=\left[\left(\f{k}{a}\right)^2-\left(\f{9}{4}H^2+\f{3}{2}H\f{\dot{\epsilon}}{\epsilon}-\f{1}{4}\f{\dot{\epsilon}^2}{\epsilon^2}\right)-\f{1}{2}\left(3\dot{H}+\f{\ddot{\epsilon}}{\epsilon}\right)\right].
\end{equation}
Modes for which the term proportional to $k$ dominates the effective frequency $\bar{\omega}_k$ tend to be adiabatic since they satisfy $|\dot{\bar{\omega}}_k/\bar{\omega}_k^2|\ll1$ during the transition.  Likewise, modes which satisfy $|\dot{\bar{\omega}}_k/\bar{\omega}_k^2|\gg1$ during the transition tend to experience a sudden transition.  We summarize these behaviors in Table \ref{tab:cases}.

\begin{table}[h]
\centering
\begin{tabular}{|c|c|}\hline
Condition&Cases\\\hline
$|\dot{\bar{\omega}}_k/\bar{\omega}_k^2|\gg1$&Sudden Transition\\
&Intermediate Modes\\
$|\dot{\bar{\omega}}_k/\bar{\omega}_k^2|\ll1$&Adiabatic\\\hline
\end{tabular}
\caption{Three cases for a given momentum mode depending on how deep inside of the horizon it is during the transition.}
\label{tab:cases}
\end{table}

From Figure \ref{fig:cases} there are clearly two distinct cases in which we may observe excited modes:\\
1.  We observe sudden/intermediate modes with an amplitude close to the maximum amplitude, in which case the fractional change in $\epsilon$ is strongly bounded (see Table \ref{tab:largest_w0}).\\
2.  We observe only intermediate modes at the low amplitude tail of the spectrum, in which case the fractional change in $\epsilon$ may have been large but the modes with a large amplitude are hidden outside of our horizon.

The second case may allow for large fractional change in $\epsilon$ compared to the bounds presented in Table \ref{tab:largest_w0}, but it requires fine tuning to ensure that the large amplitude modes are not observed.  The fine tuning becomes more concerning as the fractional change in $\epsilon$ increases, because difference in amplitude between the hidden modes and the visible modes increases dramatically.  Moreover, it is not clear we would be able to infer $\epsilon_0$ from the observation of the low amplitude tail modes.

\section{Implications for Low Multipole Scalar Power Spectrum Suppression}\label{sec:low}
Observationally there is a suppression of power for multipoles of $20\lesssim\l\lesssim30$ \cite{Planck:2015xua}.  A brief discussion of the history associated with discovering and modeling this anomaly is contained in \cite{Cicoli:2014bja}.  The observed power suppression is approximately given by
\begin{equation}
\f{\Delta_\Rd^2|_{\text{expected }20<\l<30}}{\Delta_\Rd^2|_{\text{actual }20<\l<30}}\approx \f{600}{1000}=60\%.
\end{equation}
To suppress the scalar power spectrum on large scales, one would need the scale dependence of $|\alpha_k+\beta_k|$ to suppress power for the relevant multipoles.

Based on our discussion in the previous section, there are two possibilities:\\
1. Sudden transition modes comprise the entire CMB, in which case the bounds from Table \ref{tab:largest_w0} hold.\\
2. Only the lowest $\l$ modes are excited.\\

For the first case the largest relative suppression that may be obtained is
\begin{equation}
\f{|\alpha_k+\beta_k|^2_{LB}}{|\alpha_k+\beta_k|^2_{UB}}\approx\left(\f{0.97}{1.022}\right)^2=90\%,
\end{equation}
This is an insufficient amount of suppression.

For the second case we note that the envelope of the excited spectrum typically monotonically decays from a large excitation amplitude to a smaller excitation amplitude as is depicted in Figure \ref{fig:cases}.  Since the modes between $\l\approx10$ and $\l\approx20$ do not show a power suppression to the same extent that the $20\lesssim\l\lesssim30$ modes do,  we do not think that the transitions which we have studied are good candidates for explaining the power suppression anomaly.  It may be possible to finely tune the pre-transition excitation parameter $\beta_{0,k}$ (see Figure \ref{Fig:betavx}) to obtain the desired spectrum \cite{Sriramkumar:2004pj}, but it is not obvious what mechanism could give rise to such a selected excitation.

\section{Conclusions}\label{sec:Conclusions}
We have revisited the physics of early universe transitions to slow-roll inflation.  The proper matching conditions must be used when determining the spectrum of excited fluctuations across an equation of state transition.  A careful numerical study of the problem agrees with matching $\left\{\Rd,\epsilon\dot{\Rd} \right\}$ as opposed to $\left\{\Rd,\dot{\Rd}\right\}$.  There are three regimes present in a gradual transition: modes which experience a sudden transition, adiabatic modes and modes which interpolate between those two regimes which we call intermediate modes.

If the modes comprising the visible CMB contain imprints of the transition, we have shown that the pre-transition universe must likewise be an inflationary period.  The only exception is if the cosmic variance limited modes are comprised of intermediate modes generated by a transition from a large $w_0$.  We have also argued that is it very unlikely that equation of state transitions can explain the low multipole power suppression observed in the CMB since it requires a very localized excitation in momentum space prior to the transition.

Our results state that the physics which preceded inflation is not likely to be imprinted on the observable CMB.  This is a discouraging result from the perspective of using CMB observations to gain insight into the earliest stages of our universe.  However, it is encouraging since it allows us to interpret cosmological observations in the context of inflationary cosmology without having to worry about potential ambiguities introduced by pre-inflationary physics.

\section*{Acknowledgments}
This material is based upon work supported by the National Science Foundation under Grant Number PHY-1316033 and Grant Number PHY-1521186.
%

\end{document}